\documentclass[aps,prb,twocolumn, showpacs, superscriptaddress]{revtex4-1}
\usepackage{amsmath, amsthm, amssymb,amsbsy,graphicx,times,subfigure,marvosym,color,bm,hyperref,psfrag,graphics}
\usepackage[latin1]{inputenc}
\usepackage{sidecap}
\usepackage{ulem}
\usepackage{appendix}

\begin{document}
\title{Semi-classical Approach to Quantum Spin Ice}
\author{M. P. Kwasigroch}
\affiliation{T.C.M. Group, Cavendish Laboratory, University of Cambridge, J. J. Thomson Avenue, Cambridge CB3 0HE, U.K.}
\author{B. Dou\c{c}ot}
\affiliation{LPTHE, CNRS and Universit\'{e} Pierre et Marie Curie, Sorbonne Universit\'{e}s, 75252 Paris Cedex 05, France}
\author{C. Castelnovo}
\affiliation{T.C.M. Group, Cavendish Laboratory, University of Cambridge, J. J. Thomson Avenue, Cambridge CB3 0HE, U.K.}


\begin{abstract}
We propose a semi-classical description of the low-energy properties of {\it quantum spin ice} in the strong Ising limit. Within the framework of a semiclassical, perturbative Villain expansion, that can be truncated at arbitrary order, we give an analytic and quantitative treatment of the deconfining phase. 
We find that photon-photon interactions significantly renormalise the speed of light and split the two transverse photon polarisations at intermediate wavevectors. We calculate the photon velocity and the ground state energy to first and second order in perturbation theory, respectively. The former is in good agreement with recent numerical simulations. 
\end{abstract}
\pacs{75.10.Jm, 75.10.Kt, 11.15.Ha}



\maketitle

\section{Introduction}
Classical spin-ice materials such as Ho$_2$Ti$_2$O$_7$ and Dy$_2$Ti$_2$O$_7$ contain magnetic moments that occupy sites of corner-sharing tetrahedra. Local strong crystal fields force the moments to point either in or out of the tetrahedra, motivating an effective spin$-1/2$ description. 
Dipolar interactions between the spins lead to the famous 2in-2out ice rules at low temperatures, giving a macroscopically degenerate manifold of classical spin-ice states~\cite{Bramwell2001}. Tetrahedra that violate the ice rules correspond to sources of flux of the physical magnetic field and can be identified as magnetic monopoles~\cite{Castelnovo2008,Castelnovo2012}.

The possibility of realising quantum analogues of these systems, dubbed {\it quantum spin ice}~\cite{Molavian}, in related rare-earth magnets such as Tb$_2$Ti$_2$O$_7$, Pr$_2$Sn$_2$O$_7$, Pr$_2$Zr$_2$O$_7$ and Yb$_2$Ti$_2$O$_7$, 
has attracted much attention of late. However, definitive confirmation of the discovery of quantum spin ice is yet to be found. (We point to Ref.~\onlinecite{Gingras} for an extensive survey of the theoretical and experimental progress on quantum spin ice; more recent experimental efforts include Refs.~\onlinecite{Pan2016,Tokiwa2016,Hallas2016, Kadowaki, Ong}.

This anticipated {\it quantum spin liquid} state of matter is argued to be a gapless U(1) spin liquid. Indeed, it was shown by Hermele {\it et al.}~\cite{Hermele} that quantum (virtual) perturbative processes can lead to an effective Hamiltonian that couples states in the 2in-2out manifold. Because the Hamiltonian acts within the space of states that satisfy a lattice divergenceless condition, it necessarily possesses U(1) symmetry. This gauge invariance prevents long-range order down to zero temperature and keeps the system in a quantum spin liquid phase. In the same reference, it was also demonstrated that the Hamiltonian can be mapped onto a variant of compact U(1) lattice gauge theory, the compactness arising from the discreteness of the spin$-1/2$ degrees of freedom. From lattice gauge theory literature, e.g., Ref.~\onlinecite{Kogut}, it is known that such a model exhibits two phases: a deconfining one with a gapless photon excitation, where (static) charges (magnetic monopoles) interact via Coulomb forces; and a confining phase where the photon is gapped and charges are confined. Hermele {\it et al.} argued that the underlying frustration of the spin-$1/2$ model keeps quantum spin ice in its deconfining phase. 
These results were confirmed using quantum Monte-Carlo calculations by Shannon {\it et al.}~\cite{Shannon}, who compared the dynamical structure factor as predicted by lattice gauge theory for the deconfining phase against numerical results. 

Progress has also been made away from the strong Ising limit, where the coupling between the degenerate spin-ice states is no longer small by comparison with the Ising exchange. This leads to increasing violations of the ice rules and proliferation of magnetic monopoles. Slave-boson treatments~\cite{Savary2, Savary1, Hao, Makhfudz} have studied the resulting transitions from quantum spin ice into neighbouring ordered phases through mean-field theory or phase stability arguments. In particular, in Ref.~\onlinecite{Savary1}, it was shown that a condensation of magnetic monopoles leads to a transition from the quantum spin ice phase into a phase with antiferromagnetic order. The above analytic treatments have been complemented and supported by numerical investigations, e.g., Ref.~\onlinecite{Onoda} and Ref.~\onlinecite{Banarjee}. 

For completeness, we also mention the recent theoretical work in Refs~\onlinecite{Petrova2015,Wan2016,Kourtis2016} directed at investigating the behaviour of quasiparticle (monopole) excitations in quantum spin ice. 

In the search for experimental signatures of quantum spin liquid behaviour, and in general to gain further insight on the properties of quantum spin ice systems, new theoretical perspectives can be helpful. In this Article, we propose a complementary viewpoint to that taken in Ref.~\onlinecite{Hermele}. We apply Villain's semi-classical expansion~\cite{Villain} to quantum spin ice and obtain quantitative estimates of the ground-state energy and the long-wavelength dispersion of its excitations. We find the latter to be in good quantitative agreement with the numerical results obtained in Ref.~\onlinecite{Shannon}. 
Photon-photon interactions significantly renormalise the speed of light and split the two transverse photon polarisations at intermediate wavevectors. 

In our approach, the deconfining phase and its gapless photon excitations arise naturally at large length scales, through a coarse-graining of the microscopic spin$-1/2$ degrees of freedom, analogously to how the ordered phase and its spin-wave excitations arise at large length scales in spin-$1/2$ ferromagnets. 
In light of the large-s expansion, our approach offers the advantage of being able in principle to systematically improve on the accuracy of the results by going to higher order in perturtbation. 

In Sec.~\ref{sec: large s description}, we describe in detail the effective ring-exchange Hamiltonian that was derived in Ref.~\onlinecite{Hermele} and which acts within the manifold of spin-ice states. We introduce the semi-classical perturbative large-spin expansion and we discuss, following Ref.~\onlinecite{Villain}, its surprising success in the case when $s=\frac{1}{2}$. 
A calculation of the ground state energy and dispersion is also presented here to quadratic order. 
In Sec.~\ref{sec: corrections}, we look at higher order corrections arising from photon-photon interactions. In particular, we calculate the renormalisation of the speed of light and the ground state energy. We argue that higher order terms are generally irrelevant in the RG sense. We discuss how zero-point fluctuations affect the ground state, and from this we argue that the kinematic constraints arising from the finite spin size are irrelevant at large length scales.

\section{\label{sec: large s description}
Large-S description
        }
Spin-ice materials contain magnetic moments that occupy the sites of a pyrochlore lattice. The pyrochlore lattice is a bipartite lattice of corner sharing 'up' and 'down' tetrahedra, whose centres map out a diamond lattice and the corners correspond to pyrochlore lattice sites. The crystal field forces the magnetic moments to lie along the bonds of the diamond lattice and they are well approximated by spin-1/2 degrees of freedom~\cite{Onoda Hamiltonian}. A given moment is in the $S^z= 1/2$ spin state if it points out of the 'up' tetrahedron and in the $S^z=-1/2$ state if it points into it.  Furthermore, because of strong Ising-type exchange, the low-energy manifold of spin ice obeys the following constraint
\begin{eqnarray}
\sum_{n \in {\rm tet.}} S^z_{n}=0 
\, , 
\label{constraint}
\end{eqnarray}
where the sum is taken over the four corners of a given tetrahedron. These are known as spin ice rules.

\begin{figure}
\centering
\includegraphics[width=0.9\columnwidth, angle=270]{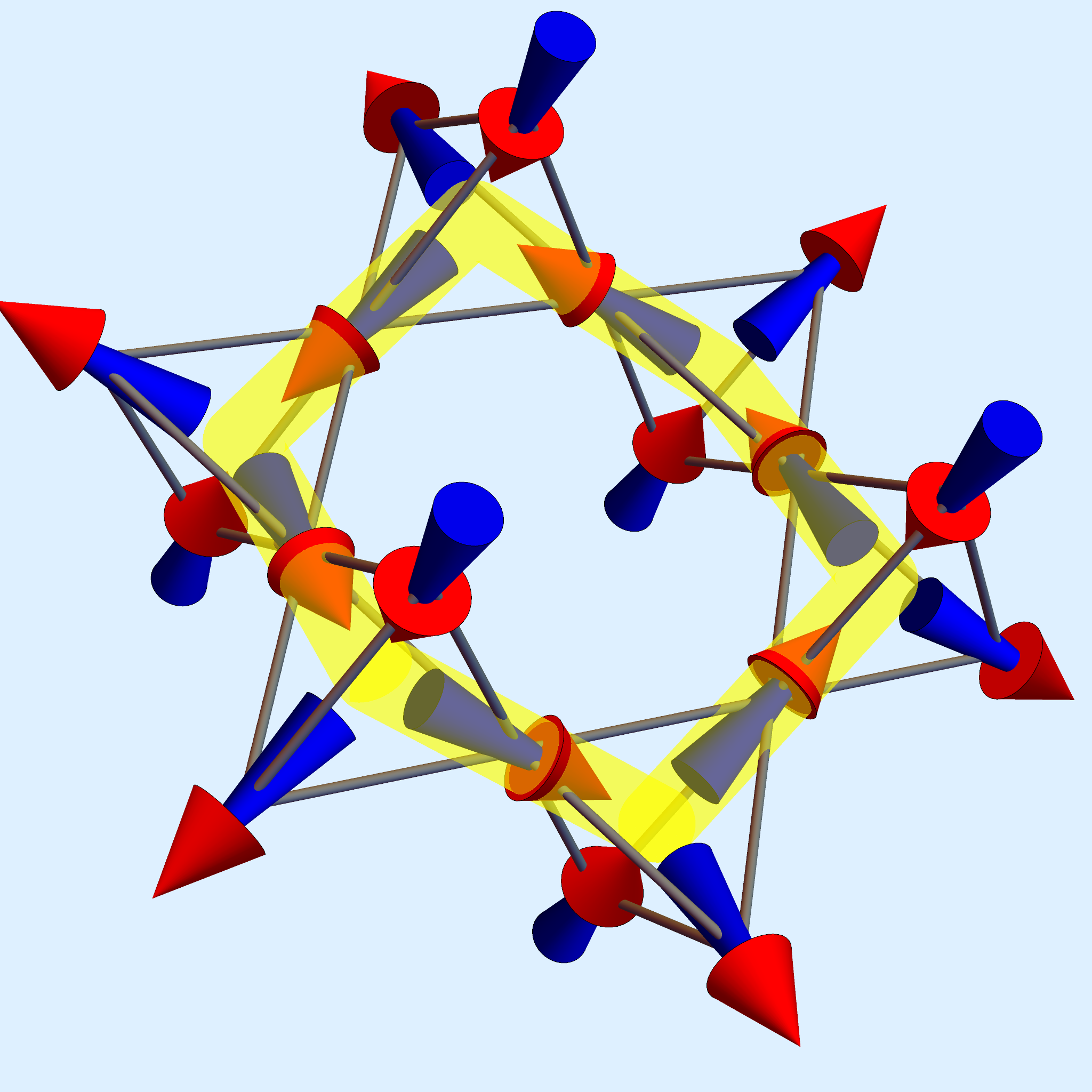}
\caption{
Hexagonal ring-exchange from the Hamiltonian in Eq.~\eqref{RingExchange}. The highlighted spins 1--6 form a single plaquette. As shown, the plaquette is flippable since adjacent spins have opposite sign of $S^z$. Note that the spins lie along lines joining the centres of adjacent tetrahedra and that, by the constraint in Eq.~\eqref{constraint}, each tetrahedron has two spins that point out of it and two spins that point in. }
\label{FigHexagon}
\end{figure}

Quantum effects lead to tunneling between the two $S^z=\pm 1/2$ spin states. The lowest order virtual process that connects two states that satisfy the spin-ice rules (and does not give rise to a trivial constant) is the hexagonal ring-exchange. This perturbative process is captured by the following effective Hamiltonian, derived in Ref.~\onlinecite{Hermele}
\begin{eqnarray}
\mathcal{H}= -g\sum_{{\rm hex.}}\left(S^+_1S^-_2S^+_3S^-_4S^+_5S^-_6 + {\rm h.c.}\right)
\, . 
\label{RingExchange}
\end{eqnarray}
Here the sum is taken over all possible hexagonal plaquettes of the pyrochlore lattice and $S^+_{n}$, $S^-_{n}$, $n=$ 1--6, are the spin-1/2 raising and lowering operators for the six spins that form a given plaquette.   Fig.~\ref{FigHexagon} highlights these spins. The Hamiltonian is a sum of terms, each of which flips all the spins around a different plaquette, assuming they are in an appropriately flippable state. 

This Hamiltonian was studied in Ref.~\onlinecite{Hermele} and Ref.~\onlinecite{Shannon}, where, by mapping the ring-exchange model to an O(2) quantum rotor representation, it was argued that the low energy physics is that of compact U(1) electrodynamics in its deconfining phase. This was later confirmed with Monte-Carlo calculations on finite systems~\cite{Shannon}, where it was found that the spin-spin correlators are chiefly governed by a linearly-dispersing photon excitation, characteristic of deconfining U(1) electrodynamics.

The O(2) rotor picture has the main drawback that no analytic, quantitative predictions can be easily made from the bare microscopic parameters. Moreover, the deconfining phase is argued to be a result of the underlying {\it frustration} -- the eigenvalues of the magnetic field $S^z$ are half-integers and hence its expectation value cannot be made to vanish, as would be energetically favourable in the confining phase where magnetic flux lines form narrow tubes. It is proposed that this frustration leads to a massive renormalisation of the bare parameters of the model such that the effective {\it long-wavelength} description is that of compact electrodynamics in its deconfining phase.

Here, we propose a complementary view that allows to shed some light on this massive renormalisation and to make quantitative predictions that were previously not accessible. With the hindsight that the effective {\it long-wavelength, low-energy} description hosts gapless, linearly-dispersing collective photon excitation, in which the original, discrete, i.e., quantum, nature of the participating spins is coarse-grained over, we propose a semi-classical large-$s$ description. 
That is, we extend our analysis of the Hamiltonian in Eq.~\eqref{RingExchange} to spins of general size $s$.

We start off with large spins and employ semi-classical approximations to obtain the effective model. The spin $s$ can then be tuned to 1/2 to obtain quantitative estimates of the relevant parameters. The success of this approach is a consequence of the RG flow to the fixed point at $\frac{1}{s}=0$, so that at small momenta $1/s$ can be treated as a small parameter, and is analogous to the success of spin-wave description in spin-1/2 quantum ferromagnets. We have implicitly assumed the absence of other fixed points -- because of the above mentioned {\it frustration}, the sytem is in the deconfining phase for all $s$ and the RG flow is controlled by the $\frac{1}{s}=0$ fixed point. Note that it is the relevant gapless, long-wavelength modes that determine the RG flow to large $s$, rather than the global broken symmetry {\it per se}. To see this, it is particularly instructive to use a spin representation that, unlike Holstein-Primakoff or Dyson-Maleev, does not rely on a broken symmetry. A particularly useful representation of this type is the Villain representation (see e.g., Ref.~\onlinecite{Nishimori}, where it is used to obtain quantitative estimates for the spin-1/2 XY model; the results agree to second order in $1/s$ with those obtained by Holstein-Primakoff).

In the Villain spin representation, the U(1) gauge symmetry of the ring-exchange Hamiltonian becomes explicit:
\begin{eqnarray}
S^+ = e^{i\phi/2}\sqrt{\tilde{s}^2-{S^z}^2}e^{i\phi/2} 
\qquad 
S^-=\left(S^+\right)^{\dagger}
\, ,
\end{eqnarray}
where $\tilde{s}=s+\frac{1}{2}$, $\phi$ and $S^z$ are canonically conjugate operators ($[\phi,S^z]=i$), and the Hilbert space is spanned by periodic eigenfunctions of $\phi$, or correspondingly, by eigenfunctions of $S^z$ with integer (or half-integer) eigenvalues. Our physical system will be further restricted to the subspace where $|S^z|\leq s$ -- this is known as the kinematic constraint. 

We introduce the variable $p=\frac{S^z}{\tilde{s}}$ and expand the Hamiltonian in Eq.~\eqref{RingExchange} to order $\tilde{s}^{-1}$, i.e. to second order in $p$ and $\phi$ (for large $s$ the quadratic term controls the fluctuations of $p$ and $\phi$, which scale as $\tilde{s}^{-\frac{1}{2}}$)

\begin{eqnarray}
\!\!\!\!\!\!\!\frac{\mathcal{H}}{g\tilde{s}^6} 
= 
-2 + 
  \sum_{\alpha \beta} 
	  \left( {\rm curl_{\alpha \beta}}\phi \right)^2
		+
		z\sum_{ij} p_{ij}^2
+
\mathcal{O}\left(\frac{1}{\tilde{s}^{2}}\right) 
\label{quadratic}
\, ,
\end{eqnarray}
where $z=6$ is the coordination number of the hexagonal plaquette, latin letters $\{i\}$ index the sites of the diamond lattice (bond midpoints $\{ij\}$ correspond to pyrochlore lattice sites on which the spins live) and the greek letters index the sites of the dual diamond lattice. Bond midpoints of the dual diamond lattice $\{\alpha\beta\}$ correspond to centres of hexagonal plaquettes of the original pyrochlore lattice (see App.~\ref{app: diag H} for an explanation of this duality). In particular, ${\rm curl_{\alpha \beta}}\phi\equiv \phi_1-\phi_2+\phi_3-\phi_4+\phi_5-\phi_6$, where 1--6 index the six spins that make up the hexagonal plaquette centred on $\alpha \beta$.

{\it Validity of the harmonic approximation.} -- By expanding the Hamiltonian in small ${\rm curl_{\alpha \beta}}\phi$ we have made an implicit approximation. Namely, we are approximating periodic wavefunctions     $\Psi_p(\{\phi_{ij}\})$ by square-integrable wavefunctions $\Psi_s(\{\phi_{ij}\})$. Because $\Psi_s(\{\phi_{ij}\})=\Psi_p(\{\phi_{ij}\})$ for $\vert {\rm curl_{\alpha \beta}}\phi \vert < \pi$ and vanishes otherwise, a smooth approximation for $\Psi_s(\{\phi_{ij}\})$ works well if most of the weight of the wavefunction is confined to the region where ${\rm curl_{\alpha \beta}}\phi\approx 0$. Note that by conjugacy, removing the periodicity in $\phi_{ij}$ is equivalent to making $p_{ij}$ continuous. The harmonic approximation works well for large $s$ because typical fluctuations $\mathcal{O}\left({\rm curl_{\alpha \beta}}\phi\right)\sim \mathcal{O}(p_{ij})\sim s^{-\frac{1}{2}}$ become small. It is surprising however that it may work well for $s=\frac{1}{2}$. However, the system is believed to be in the deconfining phase even for $s=\frac{1}{2}$, and in this phase, the long-distance properties are determined by long-wavelength, {\it gapless} degrees of freedom. As explained above, these permit us to take an average of many adjacent spins, and in this case the square-integrable wavefunctions become a good approximation of the physical periodic eigenstates of the system. Consider a single spin-1/2 aligned along the $x$ axis. Its wavefunction in the $S^z$ basis is the well-known $\Psi_{\rightarrow}(S^z)=\frac{1}{\sqrt{2}}(\delta_{S^z=\frac{1}{2}}+\delta_{S^z=-\frac{1}{2}})$, and in the $\phi$ basis it is equal to $\Psi_{\rightarrow}(\phi)=\frac{1}{\sqrt{\pi}}\cos(\frac{\phi}{2})$. A square-integrable approximation for $\Psi_{\rightarrow}(\phi)$ or a continuum approximation for $\Psi_{\rightarrow}(S^z)$ at this stage would be rather poor. However, if we have a large number $N$ of spins aligned along the $x$ axis and we are interested in the average $\langle\phi\rangle$ or average $\langle S^z\rangle$ of this ensemble (coarse-graining), then by the central limit theorem, a Gaussian approximation for these will work very well: $\Psi_{\rightarrow}(\langle\phi\rangle)=e^{-3N\langle\phi\rangle^2/2\pi(\pi^2-6)}$ and $\Psi_{\rightarrow}(\langle S^z\rangle)=e^{-2N\langle S^z\rangle^2}$, regardless of the underlying distribution or its discreteness. The only parameters that the individual spins provide are the average and variance of the relevant variable, which is set by $s$. This is the essence of why the harmonic approximation works so well even for $s=\frac{1}{2}$, when long-wavelength modes that allow for coarse graining are the relevant degrees of freedom in the system.

The quadratic part of the Hamiltonian is diagonal in the appropriate basis (details of the required transformation, which follows Ref.~\onlinecite{Shannon}, can be found in App.~\ref{app: diag H}): 
\begin{eqnarray}
\!\!\! 
\frac{\mathcal{H}_0}{g\tilde{s}^6} 
\! = \!\!\! 
\sum_{\lambda, \mathbf{k}\in{\rm BZ}}\left[\xi_{\lambda}^2 (\mathbf{k}) \phi_{\lambda}( \mathbf{k})\phi_{\lambda}(- \mathbf{k})+zp_{\lambda}( \mathbf{k})p_{\lambda}(- \mathbf{k})\right]
, 
\label{quadratic in k} 
\end{eqnarray}
where $\lambda=1,2,3,4$ indexes the four normal mode branches, $k$ is summed over the first Brillouin zone of the fcc lattice, and $\xi_{\lambda}( \mathbf{k})$ are given by
\begin{eqnarray}
&&\xi_{\lambda=1,2}( \mathbf{k})
=
\pm\sqrt{2}\sqrt{\sum_{\mu \nu}\sin^2 \left( \mathbf{k} \cdot  \mathbf{\Delta}_{\mu \nu}\right)}
\, ,
\nonumber\\
&&
\xi_{\lambda =3,4}( \mathbf{k}) 
=0 
\, , 
\label{eq: xi}
\end{eqnarray}
and the vectors $ \mathbf{\Delta}^{\mu \nu}$ are given in the appendix. The conjugate operators satisfy the canonical commutation relations 
$
[ \phi_{\lambda}( \mathbf{k}),p_{\lambda '}^{\dagger}( \mathbf{q}) ] 
= 
\frac{1}{\tilde{s}}\delta_{ \mathbf{k}, \mathbf{q}}\delta_{\lambda,\lambda '}
$. 
There are two divergenceless modes $\lambda=1,2$, which correspond to the two polarisations of the photon and become the two transverse modes in the continuum limit, and there are two divergenceful modes $\lambda=3,4$ which give rise to the longitudinal mode in the continuum limit. 
The spin-ice rules enforce zero lattice divergence on $p_{ij}$ so that the divergenceful modes vanish identically, 
\begin{eqnarray}
p_{\lambda=3}( \mathbf{k})=p_{\lambda=4}( \mathbf{k})=0 \:\:\: {\rm for\: all\:}  \mathbf{k} 
\, . 
\label{eq: divless constraint}
\end{eqnarray}
Because of this constraint, the divergenceful modes do not enter the Hamiltonian. These degrees of freedom span the $2N_s$-dimensional space of constants of motion (where $N_s$ is the number of $\mathbf{k}$ vectors in the Brillouin zone of the fcc lattice), i.e., the above operators commute with the Hamiltonian, and are in a one-to-one correspondence with the tetrahedron charges $\sum_{i \in {\rm tet.}} p_{i}$, which span the same space. Because $p_{\lambda=3}(\mathbf{k})$ and $p_{\lambda=4}(\mathbf{k})$ are linear combinations of the tetrahedron charges, they must vanish in their absence. 
Note that, by the uncertainty principle, the fluctuations in the divergenceful part of $\phi_{ij}$, i.e., $\phi_{\lambda=3}(\mathbf{k})$ and $\phi_{\lambda=4}(\mathbf{k})$, are unbounded and correspond to the U(1) gauge freedom of the Hamiltonian: $\phi_{ij}\rightarrow \phi_{ij}+\chi_i-\chi_j$. 

For the two divergenceless modes $\lambda=1,2$ that remain in the Hamiltonian, we introduce bosonic creation/annihilation operators 
\begin{eqnarray}
\phi_{\lambda}( \mathbf{k}) 
&=& 
\sqrt{\frac{\omega( \mathbf{k})}{2\tilde{s}}}
\left(
  a_{\lambda}^{\dagger}( \mathbf{k})+a_\lambda(- \mathbf{k})
\right) 
\, ,
\nonumber\\
p_{\lambda}( \mathbf{k})
&=&
\frac{i}{\sqrt{2\tilde{s}\omega( \mathbf{k})}}
\left(
  a_{\lambda}^{\dagger}( \mathbf{k})-a_\lambda(- \mathbf{k})
\right)
\, , 
\label{diagonalisation}
\end{eqnarray}
where the bosonic operators obey the usual commutation relations, in particular $
[ a_{\lambda}( \mathbf{k}),a_{\lambda '}^{\dagger}( \mathbf{q}) ] 
= 
\delta_{ \mathbf{k}, \mathbf{q}}\delta_{\lambda,\lambda '}
$. 
If we make the choice
\begin{eqnarray}
\omega(\mathbf{k})=\sqrt{z}/|\xi_{\lambda = 1}(\mathbf{k})|,
\end{eqnarray}
the Hamiltonian becomes diagonal in the above basis
\begin{eqnarray}
\mathcal{H}_0 
=
\sum_{\mathbf{k}\in {\rm BZ},\lambda=1,2}\epsilon(\mathbf{k})\left(
  a_{\lambda}^{\dagger}(\mathbf{k})a_\lambda(\mathbf{k})+\frac{1}{2}
\right) 
\, , 
\label{eq: H0}
\end{eqnarray}
where the elementary spin-wave excitations $a_{\lambda}(\mathbf{k})$ are 'photon' like, i.e. they are gapless, linearly dispersing modes with two polarisations $\lambda=1,2$. The energy dispersion 
\begin{eqnarray}
&&\frac{\epsilon(\mathbf{k})}{g s^5}
=
\frac{2z}{\omega(\mathbf{k})}
+
\mathcal{O}\left(\frac{1}{s}\right)
\quad 
\overset{|\mathbf{k}|a_o \ll 1 }{\rightarrow} 
\quad 
c |\mathbf{k}|
\, , 
\nonumber\\
&&c\approx 0.15 g a_0 \qquad {\rm for} \quad s=\frac{1}{2}
\, .
\end{eqnarray}
This is substantially different from the Monte Carlo estimate of Ref.~\onlinecite{Shannon}. We will see that there is a sizeable Hartree-Fock correction coming from higher order terms in the Hamiltonian (Eq.~\eqref{RingExchange}) which we believe is chiefly responsible for this discrepancy. The total energy $E$, to order $1/s$ is given by
\begin{eqnarray}
\frac{E}{4N_s g s^6}&=&-2\frac{\tilde{s}^6}{s^6}+ \frac{\sqrt{z}\tilde{s}^5}{4N_s s^6}\sum_{\mathbf{k}, \lambda}|\xi_{\lambda}(\mathbf{k})| + \mathcal{O}\left(\frac{\tilde{s}^4}{s^6}\right)
\nonumber\\
&=&-2+ \frac{1}{s}\left[-6+\frac{\sqrt{z}}{4N_s}\sum_{\mathbf{k}, \lambda }|\xi_{\lambda}(\mathbf{k})|\right] +\mathcal{O}\left(\frac{1}{s^2}\right)
\nonumber\\
&=&-2+\frac{A_1}{s}+\mathcal{O}\left(\frac{1}{s^2}\right) 
\, , 
\label{ZPF}
\end{eqnarray}
where $A_1\approx -1.820$ and the two corrections at order $1/s$ arise from the smearing of the spin length and zero-point fluctuations, respectively. $N_s$ is the number of $\mathbf{k}$-vectors in the first Brillouin zone of the fcc lattice and $4N_s$ is the number of pyrochlore lattice sites. 

It is instructive to compare this result with that for the square spin-$1/2$ XY model from Ref.~\onlinecite{Nishimori}, 
\begin{eqnarray}
E_{\rm XY}/Ns^2=-2-\frac{0.084}{s}+\mathcal{O}\left(\frac{1}{s^2}\right) 
\, ,
\end{eqnarray}
where the same Villain spin representation was used. The coefficient of the $1/s$ term is much larger in our case and this can be explained by the fact that there are six spins participating in the ring-exchange interaction as compared to two in the easy-plane ferromagnetic exchange of the XY model. Zero point fluctuations contribute at second order in the expansion of the Hamiltonian around the classical saddle pont. Writing each spin as the sum of its classical expectation value (gauging out unbounded longitudinal fluctuations in the case of ring exchange -- see App.~\ref{app: gMFT parameter}) and a small fluctuation $\delta S$, we see that there will be $^6C_2 = 15$ quadratic terms for the ring exchange (for each plaquette) and $^2C_2 = 1$ such terms for the XY model (for each bond). Assuming each quadratic term gives a separate contribution $(\delta S/S)^2$ that scales as $1/s$ (largely set by infrared fluctuations of the spin phase which are insensitive to the microscopic model), we find that the ratio of zero-point fluctuations for the two models scales roughly as $^6C_2/^2C_2=15$, in agreement with the results above. 

We close by noticing that this quadratic analysis allows also to compute the zero-point fluctuations of the gauge mean-field, as defined, for example, by the slave boson mapping of Ref.~\onlinecite{Savary1} (see App.~\ref{app: gMFT parameter}). We find a $\sim 4\%$ reduction of the gauge mean-field, suggesting that in the strong Ising limit $g \rightarrow 0$, there are only small corrections to gauge mean-field theory from zero-point fluctuations.

\section{\label{sec: corrections}
Perturbative corrections: Spin-Wave Interactions
        }
We now consider perturbative corrections to the energy of the system, at order $1/s^2$ in $E/gs^6$, arising from normal ordering of quartic terms in the Hamiltonian in Eq.~\eqref{RingExchange}, i.e., Hartree-Fock corrections. We shall not consider here self-energy terms (which come in at order higher than $1/s^2$), although our perturbative expansion can be straightforwardly extended to compute them. 
\subsection{Hartree-Fock corrections}
We first note that there are no cubic terms in the expansion of the Hamiltonian in Eq.~\eqref{RingExchange}. Collecting all quartic terms, we get 
\begin{eqnarray}
&&\frac{\mathcal{H}_I}{g\tilde{s}^6} 
= 
-\sum_{\alpha \beta}\Big[\frac{2}{4!}\left({\rm curl}_{\alpha \beta}\phi\right)^4-\frac{1}{2}\sum_{ij \in \alpha\beta}p_{ij}^4
\label{quartic}
\\ 
&&+\frac{1}{4}\left({\rm curl}_{\alpha \beta}\phi\right)\left(\sum_{ij \in \alpha\beta}p_{ij}^2\right)\left({\rm curl}_{\alpha \beta}\phi\right)+\frac{1}{4}\sum_{ij,kl \in \alpha \beta}p_{ij}^2p_{kl}^2
\nonumber\\
&&+\frac{1}{8}\left({\rm curl}_{\alpha \beta}\phi\right)^2\left(\sum_{ij \in \alpha\beta}p_{ij}^2\right)+\frac{1}{8}\left(\sum_{ij \in \alpha\beta}p_{ij}^2\right)\left({\rm curl}_{\alpha \beta}\phi\right)^2\Big]
\, ,
\nonumber
\end{eqnarray}
where $ij \in \alpha\beta$ signifies that the sum is taken over the sites $ij$ that belong to the plaquette $\alpha \beta$. Note that the divergenceful ($\lambda=3,4$) $\phi_{ij}$ modes do not enter the Hamiltonian at all orders, and that the divergenceful $p_{ij}$ modes vanish by the constraint.  

The Hartree ground state energy correction is given by:
\begin{eqnarray}
\langle {\rm g.s.}|\mathcal{H}_I |{\rm g.s.}\rangle,
\end{eqnarray}

where $|{\rm g.s.}\rangle$ is the ground state of the quadratic Hamiltonian $\mathcal{H}_0$. The correction is equal to the constant remaining after $\mathcal{H}_I$ is normal-ordered. Details are given in App~\ref{app: normal ordering}. To order $1/s^2$ the ground state energy can then be written as
\begin{eqnarray}
\frac{E}{4N_sgs^6}&=&-2+\frac{A_1}{s}+\frac{A_2}{s^2} + \mathcal{O}\left(\frac{1}{s^3}\right) 
\, , 
\nonumber\\
E&\approx&-0.138 N_pg \qquad {\rm for} \quad s = \frac{1}{2} 
\, ,
\end{eqnarray}
where $A_2=-0.793$ and $N_p=4N_s$ is the total number of hexagonal plaquettes or pyrochlore lattice sites.

\begin{figure}
\centering
\includegraphics[width=0.9\columnwidth, angle=270]{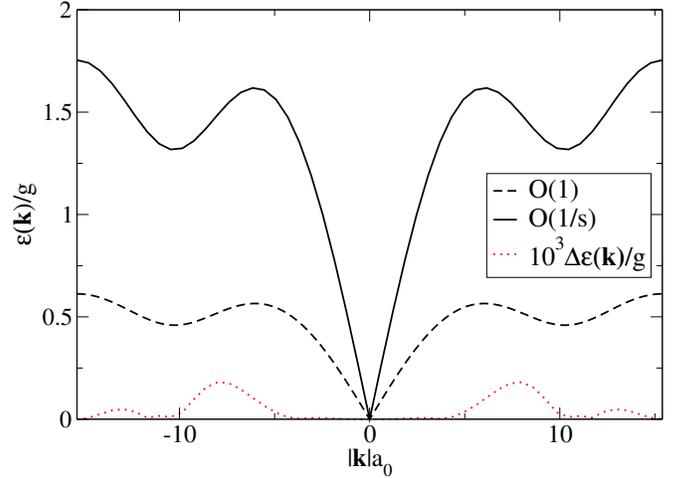}
\caption{The dispersion for the case $s=\frac{1}{2}$ in the direction [013] of the bcc reciprocal lattice. The dashed line shows the zeroth order dispersion ($\epsilon(\mathbf{k})/gs^5$ is given to zeroth order in $1/s$). The solid line includes the first order Hartree-Fock corrections, which can be seen to increase the speed of light. 
It also introduces a small splitting $\Delta\epsilon(\mathbf{k})$ of the modes away from $\mathbf{k} = 0$, which is visible on the scale of the plot {\it only} if appropriately magnified (dotted line).}
\label{dispersion}
\end{figure}

Considering the Hartree correction to the excitation spectrum, we find that $\mathcal{H}_I$ mixes the degenerate spin-wave modes, 
\begin{eqnarray}
\langle {\rm g.s.}|a_{\lambda}(\mathbf{k})\mathcal{H}_I a^{\dagger}_{\lambda'}(\mathbf{k})|{\rm g.s.}\rangle \neq 0,
\end{eqnarray}
even for $\lambda \neq \lambda '$. This is because of terms $p_{ij}^2p_{kl}^2$ in $\mathcal{H}_I$, which couple electric fields at different lattice points. The resulting splitting in the spin-wave spectrum only appears at $|\mathbf{k}|a_0 \sim 1 $ and vanishes in the continuum limit $|\mathbf{k}|a_0\rightarrow 0$, i.e. it does not lift the degeneracy of the gapless photon. Fig.~\ref{dispersion} shows the Hartree-renormalised dispersion across the Brillouin zone (details of the calculation can be found in App~\ref{app: normal ordering}).
 To order $s^{-1}$, the renormalised speed of light is given by
\begin{eqnarray}
&&\frac{c}{gs^5} 
= 
2a_0\sqrt{z}\left(1+ \frac{0.846}{s}\right)+\mathcal{O}(s^{-2}),\nonumber\\
&&c\approx 0.41 ga_0 \qquad {\rm for} \quad s=\frac{1}{2} 
\, ,
\end{eqnarray}
which is now much closer to the numerical value of $\left( 0.6 \pm 0.1 \right) ga_0$ computed in Ref.~\onlinecite{Shannon}.

\subsection{RG considerations, spin-wave damping and the ground state wavefunction}
We now shed a bit more light on the success of the large $s$ expansion by discussing the relevance of the successive terms in the Hamiltonian in the long-wavelength limit. In the path-integral description of the Hamiltonian in Eq.~\eqref{quadratic}, the continuum-limit action is of the following form (without loss of generality $g\tilde{s}^6$ has been set to one)
\begin{eqnarray}
\mathcal{S}&=& \int d^d\mathbf{r} \int d\tau \tilde{s}\Big[\left( \mathbf{\nabla} \times {\Phi} (\mathbf{r},\tau) \right)^2+ \dot{\Phi}(\mathbf{r},\tau)^2\nonumber\\
&&+\mathcal{O}\left(\left(\mathbf{\nabla}\times\Phi\right)^4 ,\dot{\Phi}^2\left(\nabla\times\Phi\right)^2,\dot{\Phi}^4\right)\Big]
\end{eqnarray}
where $\Phi(\mathbf{r},\tau)$ is a vector field and the quartic and higher order terms are included in $\mathcal{O}\left(\left(\nabla\times\Phi\right)^4,\dot{\Phi}^2\left(\nabla\times\Phi\right)^2,\dot{\Phi}^4\right)$, which is typical of the scaling properties of all the quartic terms in the Hamiltonian (see Eq.~\eqref{quartic}). Note that we have rescaled the imaginary time $\tau\rightarrow \tilde{s}\tau$ and perturbatively integrated out the massive modes. Naive RG scaling proceeds as follows:
\begin{eqnarray}
\mathbf{r},\tau &\rightarrow& b\mathbf{r}, b\tau \nonumber\\
\Phi &\rightarrow& b^{\frac{1-d}{2}}\tilde{s}^{-\frac{1}{2}}\Phi,\nonumber\\
\int d\tau \int d^d\mathbf{r} \left(\nabla \times \Phi\right)^4 &\rightarrow& \frac{b^{-d-1}}{\tilde{s}} \int d\tau \int d^dr \left(\nabla \times \Phi\right)^4 
\, , 
\nonumber\\
\end{eqnarray}
for $b>1$.
The Gaussian term dominates in all dimensions and higher order $\Phi$ terms are irrelevant. The above scaling also shows that higher order terms in $\Phi$ correspond to increasing orders in the $1/s$ expansion. Naive scaling therefore leads to the conclusion that the $1/s$ expansion provides a good description of long-wavelength physics.

From the irrelevance of the quartic term, one also expects the damping of a photon above the ground state to be small. In fact, the decay of a single photon intro three photons (caused by $\mathcal{H}_I$) vanishes exactly in the relativistic part of the spectrum because of kinematic constraints.

We have shown that dynamical spin-wave interactions are irrelevant for low energy physics. The Gaussian term dominates the action, which corresponds to neglecting {\it normal ordered} terms in the quartic contribution Eq.~\eqref{quartic}, and higher order. The renormalised quadratic Hamiltonian (i.e., after Hartree Fock corrections have been added to it) should give us good estimates for long-distance correlators.

Now that we have satisfied ourselves that dynamical spin-wave interactions are negligible, one can also try to address the issue of kinematic spin-wave interactions arising from the fact that $|S^z|\leq s$. Here, it proves very instructive to consider the overlap of a particular spin configuration with the ground state
\begin{eqnarray}
\langle 
  \{ S^z_{ij} \} \vert {\rm g.s.} 
\rangle \propto 
\left(
  \prod_{\mathbf{k},\lambda = 3,4}
	\delta_{S^z_{\lambda}(\mathbf{k})}
\right) 
e^{-\sum_{\mathbf{k},\lambda=1,2 } \frac{\omega(\mathbf{k})}{\tilde{s}}  S^z_{\lambda}(\mathbf{k})^2}
\, , 
\nonumber\\ 
\label{gs}
\end{eqnarray} 
where  $\vert \{S^z_{ij}\} \rangle$ is a particular eigenstate of $S^z_{ij}$. 
Firstly, in the limit $s\rightarrow \infty$, zero-point fluctuations disappear and the ground state becomes an unweighted (for $S^z_{ij} \ll s$) superposition of all states that satisfy the spin ice rules, i.e. those where $S^z_{\lambda=3}(\mathbf{k})=S^z_{\lambda=4}(\mathbf{k})=0$ for all $\mathbf{k}$. For $s=\frac{1}{2}$ this would correspond to the RK state. As we make $s$ finite, zero-point fluctuations have the strongest effect on the low lying $\mathbf{k}$ states, where the weights vanish non-perturbatively because $\omega(\mathbf{k}) \propto \frac{1}{\mathbf{k}}$. It is this divergence of $\omega(\mathbf{k})$ which dominates the long-wavelength physics and is for instance responsible for the disappearance of pinch-points as $s$ becomes finite and we move away from the RK state. One could therefore exclude the high Fourier components of $S^z(\mathbf{k})$ from the weights in Eq.~\eqref{gs}, since the effect on low-energy physics is negligible -- it is for these high Fourier components that the kinematic bound on the microscopic spins is also highly relevant. For low $\mathbf{k}$ components, on the other hand, the spins can be coarse-grained into large effective spins of typical size $\sim s/(|\mathbf{k}|a)^3$ and the bound becomes irrelevant.

\section{Conclusion}

We have developed a semi-classical description that accurately captures the properties of quantum spin ice at large length scales, and allows for a systematic, perturbative expansion which in principle can be truncated at arbitrary order. 
In particular, we have computed the speed of light to first order in the expansion parameter (the inverse of the spin size), and the ground state energy to second order. Our results are in good quantitative agreement with recent numerical calculations in Ref.~\onlinecite{Shannon}. We find that Hartree-Fock corrections, due to photon-photon interactions, that go beyond the quadratic U(1) lattice gauge theory, significantly renormalise the speed of light and give rise to a small splitting in the energy of the two photon modes at intermediate wavevectors. 

We offered some {\it a posteriori} justification for the semiclassical expansion in the case $s=\frac{1}{2}$ and argued that higher order terms in the expansion are irrelevant in the usual RG sense. We used some of Villain's original arguments~\cite{Villain} to also argue that square-integrable wavefunctions can provide an accurate description of the long-distance properties of the deconfining phase, which are determined by long-wavelength fluctuations of the spins. Further, we have explicitly showed how zero-point fluctuations modify the classical ground state, which is an unweighted superposition of spin-ice states. From this, we argued that the kinematic constraint on the spin size is irrelevant for the long-distance properties of quantum spin ice. 

We have also looked at how zero-point fluctuations modify the gauge mean-field theory of quantum spin ice~\cite{Savary1}. We have found, with the optimal choice of gauge (U(1) gauge freedom would otherwise prevent spontaneous symmetry breaking of the gauge field), that there is only a small $\sim 4\%$ reduction of the mean-field value at the quadratic level of the semi-classical expansion.

\acknowledgments{
This work was supported in part by EPSRC Grant No. EP/K028960/1 and by the 
EPSRC NetworkPlus on ``Emergence and Physics far from Equilibrium''. 
We gratefully acknowledge discussions with G.~Goldstein.
}

\appendix

\section{\label{app: diag H}
Diagonalising the Quadratic Hamiltonian
        }
\begin{figure}
\centering
\includegraphics[width=0.9\columnwidth, angle=0]{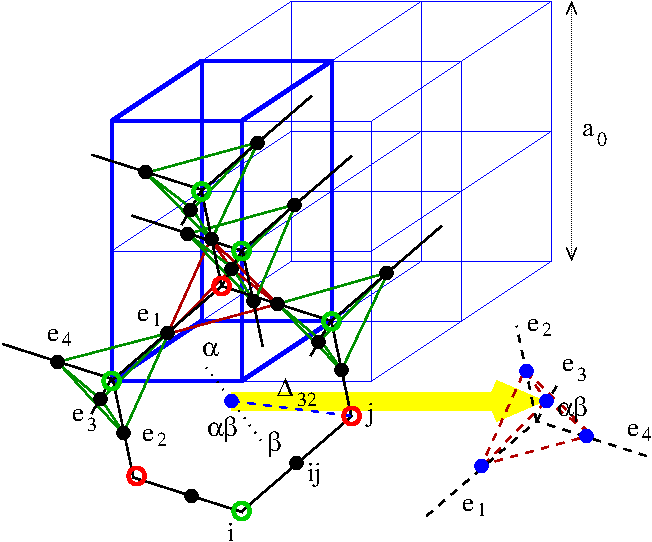}
\caption{\label{FigDual}
The pyrochlore lattice (sites shown by filled black circles) is composed of corner sharing 'up' (in red) and 'down' (in green) tetrahedra. Open circles mark their centres and these map out a diamond lattice. The fcc lattice mapped out by the centres of 'down' tetrahedra is also shown ($a_0$ is the length of the cubic unit cell). A single plaquette centred on site $\alpha\beta$ of the dual pyrochlore lattice is also drawn. The plaquette chosen is located at the $\mathbf{e_3}$ corner of one of the tetrahedra (dashed lines) of the dual pyrochlore lattice (drawn displaced by the thick yellow arrow, for convenience). The normal to the plaquette plane is along $\mathbf{e_3}$ and the vectors $\pm\mathbf{\Delta}_{3\mu}$, $\mu\neq 3$, give the positions of the plaquette vertices relative to its centre at $\alpha\beta$.} 
\end{figure}

Fig.~\ref{FigDual} summarises the geometry of the pyrochlore lattice on which the spins live, and the dual pyrochlore lattice mapped out by centres of the hexagonal plaquettes.

The pyrochlore lattice is not a Bravais lattice and hence does not have a well defined reciprocal lattice space. However, it can be decomposed into four offset fcc lattices. Considering a single 'up' tetrahedron, the position vectors of the four neighbouring 'down' tetrahedra, relative to its centre, are given by:
\begin{eqnarray}
&&\mathbf{e_1}= \frac{a_0}{4}(1,1,1), \:\: \mathbf{e}_2=\frac{a_0}{4}(1,-1,-1)\nonumber\\
&&\mathbf{e}_3=\frac{a_0}{4}(-1,1,-1), \:\: \mathbf{e}_4=\frac{a_0}{4}(-1,-1,1),
\end{eqnarray}
where $a_0$ is the side length of the fcc cubic unit cell.
The corners of the 'up' tetrahedron correspond to the midpoints of these position vectors. Each corner is a pyrochlore lattice site and is identified by one of the above vectors $\mathbf{e}_{\mu}$, where $\mu=1,2,3,4$. Corners of all 'up' tetrahedra with the same $\mu$ map out a single fcc lattice. The superposition of the four fcc lattices, one for each value of $\mu$, gives us the pyrochlore lattice.

Within the above picture, each pyrochlore lattice site can be identified by an index $\mu$, which tells us which fcc lattice it belongs to, and a position vector on that lattice. This is reflected by the following change of notation:
\begin{eqnarray}
\phi_{ij}\rightarrow \phi_{\mu} (\mathbf{r}_i+\mathbf{e}_{\mu}/2),\nonumber\\
p_{ij}\rightarrow p_{\mu} (\mathbf{r}_i+\mathbf{e}_{\mu}/2).
\end{eqnarray}
 Here $\mu$ identifies the fcc lattice to which the site $ij$ belongs ($\mathbf{e}_{\mu}=\mathbf{r}_{j}-\mathbf{r}_{i}$, where $\mathbf{r}_{j}$ and $\mathbf{r}_{i}$ are the position vectors of the 'down' and 'up' tetrahedra touching at the site $ij$) and $(\mathbf{r}_i+\mathbf{e}_{\mu}/2)$ is its position vector on that lattice.

Following Ref.~\onlinecite{Shannon}, we can now concisely express the lattice curl. We first introduce a set of vectors $\pm\mathbf{\Delta}_{\mu\nu}$ which give the positions of plaquette vertices relative to site $\alpha\beta$ of the dual pyrochlore lattice on which the plaquette is centred:
\begin{eqnarray}
\mathbf{\Delta}_{\mu \nu} \equiv \frac{a_0}{\sqrt{8}}\frac{\mathbf{e}_{\mu}\times \mathbf{e}_{\nu}}{|\mathbf{e_{\mu}}\times \mathbf{e_{\nu}}|} 
\, . 
\end{eqnarray}
The index $\mu$ identifies which of the four {\it dual} fcc lattices the site $\alpha\beta$ belongs to  ($\mathbf{e}_{\mu}=\mathbf{r}_{\beta}-\mathbf{r}_{\alpha}$, where $\mathbf{r}_{\beta}$ and $\mathbf{r}_{\alpha}$ are the position vectors of the dual 'down' and 'up' tetrahedra touching at the site $\alpha\beta$), and the index $\nu$ identifies which of the four fcc lattices the relevant plaquette vertex belongs to.
The lattice curl can now be written as
\begin{eqnarray}
{\rm curl}_{\alpha \beta}\phi&\equiv& \sum_{\nu\neq\mu,\pm} \pm\phi_{\nu}(\mathbf{r}_{\alpha}+\mathbf{e}_{\mu}/2 \pm\mathbf{\Delta}_{\mu \nu})
\, . 
\label{eq: curl in real space}
\end{eqnarray} 

We express the operators in Fourier space (introducing the Fourier transformed $\hat{\phi}_{\nu}(\mathbf{k})$ and $\hat{p}_\nu(\mathbf{k})$): 
\begin{eqnarray}
\phi_{\mu} (\mathbf{r}_i+\mathbf{e}_{\mu}/2)
&=&
\sum_{\mathbf{k}\in {\rm BZ}}\hat{\phi}_{\mu}(\mathbf{k})e^{-i\mathbf{k}\cdot(\mathbf{r}_{i}+\mathbf{e}_{\mu}/2)}
\nonumber \\ 
 p_{\mu} (\mathbf{r}_i+\mathbf{e}_{\mu}/2)
&=& 
\sum_{\mathbf{k}\in {\rm BZ}}\hat{p}_\mu(\mathbf{k})e^{-i\mathbf{k}\cdot(\mathbf{r}_{i}+\mathbf{e}_{\mu}/2)}
\, , 
\end{eqnarray} 
and the lattice curl becomes
\begin{eqnarray}
{\rm curl}_{\alpha \beta}\phi
&=&
\frac{1}{\sqrt{N_s}}
\sum_{\mathbf{k}\in {\rm BZ},\nu} 
Z_{\mu \nu}(\mathbf{k}) 
e^{-i\mathbf{k}\cdot(\mathbf{r}_{\alpha}+\mathbf{e}_{\mu}/2)}
\hat{\phi}_{\nu}(\mathbf{k}),
\nonumber\\
\label{eq: curl in k}
\end{eqnarray}
where  $Z_{\mu \nu}(\mathbf{k})=-2i\sin\left(\mathbf{k}\cdot\mathbf{\Delta}_{\mu \nu}\right)$, $N_s$ is the number of fcc lattice sites, and $\mathbf{k}$ vectors are summed over the Brillouin zone of the fcc lattice (with periodic boundary conditions).
We can then diagonalise the quadratic Hamiltonian in Eq.~\eqref{quadratic} by a unitary transformation to the eigenbasis of the $4\times 4$ matrix $Z_{\mu \nu}(\mathbf{k})$:
\begin{eqnarray}
\phi_{\lambda}(\mathbf{k}) 
&=& 
\sum_{\mu} 
U_{\lambda \mu}(\mathbf{k}) \hat{\phi}_{\mu}(\mathbf{k}) 
\nonumber \\ 
&=& 
\frac{1}{\sqrt{N_s}} \sum_{\mu, \mathbf{r}\in {\rm fcc}} 
U_{\lambda \mu}(\mathbf{k})\phi_{\mu}(\mathbf{r}+\mathbf{e}_{\mu}/2) e^{i\mathbf{k}\cdot(\mathbf{r}+\mathbf{e}_{\mu}/2)} 
\, , 
\nonumber \\ 
p_\lambda(\mathbf{k})
&=& 
\sum_{\mu} 
U_{\lambda \mu}(\mathbf{k}) \hat{p}_{\mu}(\mathbf{k}) 
\nonumber \\ 
&=&
\frac{1}{\sqrt{N_s}} \sum_{\mu, \mathbf{r}\in {\rm fcc}} 
U_{\lambda \mu}(\mathbf{k})p_{\mu}(\mathbf{r}+\mathbf{e}_{\mu}/2) e^{i\mathbf{k}\cdot(\mathbf{r}+\mathbf{e}_{\mu}/2)} 
\, , 
\nonumber\\
\label{basis}
\end{eqnarray}
where $U_{\lambda \mu}(\mathbf{k})$ is a unitary matrix chosen to diagonalise the hermitian matrix $Z_{\mu \nu}(\mathbf{k})$ and $\lambda=1,2,3,4$ indexes its eigenbasis. 
Note that the transformation from $\phi_{ij}$ and $p_{ij}$ to $\phi_{\lambda}(\mathbf{k})$ and $p_\lambda(\mathbf{k})$ is unitary and therefore preserves the canonical commutation relations. 
($\phi_{\lambda}(\mathbf{k})$ and $p_{\lambda}(\mathbf{k})$ are normal modes, not to be confused with the Fourier transforms $\hat{\phi}_\nu(\mathbf{k})$ and $\hat{p}_\nu(\mathbf{k})$, hence the use of a `hat' notation for the latter.)

For convenience we also give the inverse transformations. Firstly, we notice that $p_{3}(\mathbf{k}) = 0$ and $p_{4}(\mathbf{k}) = 0$ by the divergenceless constraint (see Eq.~\eqref{eq: divless constraint}), and therefore: 
\begin{eqnarray}
p_{\mu}(\mathbf{r}+\mathbf{e}_{\mu}/2)
&=&
\frac{1}{\sqrt{N_s}} \sum_{\lambda = 1,2, \, \mathbf{k}\in {\rm BZ}} 
\!\!\!\!\!\!
U^{\dagger}_{\mu\lambda}(\mathbf{k})e^{-i\mathbf{k}\cdot(\mathbf{r}+\mathbf{e}_{\mu}/2)}p_{\lambda}(\mathbf{k}) \, ,
\nonumber\\ 
\label{eq: p inverse}
\end{eqnarray}
where $U^{\dagger}(\mathbf{k})$ is the hermitian conjugate of the matrix $U(\mathbf{k})$. 
For $\phi$ instead, all modes are needed: 
\begin{eqnarray}
\phi_{\mu}(\mathbf{r}+\mathbf{e}_{\mu}/2)
&=&
\frac{1}{\sqrt{N_s}} \sum_{\lambda, \mathbf{k}\in {\rm BZ}} U^{\dagger}_{\mu\lambda}(\mathbf{k})e^{-i\mathbf{k}\cdot(\mathbf{r}+\mathbf{e}_{\mu}/2)}\phi_{\lambda}(\mathbf{k}) 
\, . 
\nonumber\\ 
\label{eq: phi inverse}
\end{eqnarray}
However, the $\phi$ terms enter the Hamiltonian only in the form of lattice curl, and from Eq.~\eqref{eq: curl in k} we see that 
\begin{eqnarray}
{\rm curl}_{\alpha\beta}(\phi)
&=&
\frac{1}{\sqrt{N_s}}
\sum_{\mathbf{k} \in {\rm BZ}}\sum_{\nu,\lambda} 
Z_{\mu\nu}(\mathbf{k})U^{\dagger}_{\nu\lambda}(\mathbf{k})
\nonumber\\
&\times&
e^{-i\mathbf{k}\cdot(\mathbf{r}_{\alpha}+\mathbf{e}_{\mu}/2)}
\phi_{\lambda}(\mathbf{k})
\nonumber\\
&=&
\frac{1}{\sqrt{N_s}}
\sum_{\lambda,\mathbf{k} \in {\rm BZ}}
\xi_{\lambda}(\mathbf{k})U^{\dagger}_{\mu\lambda}(\mathbf{k})
e^{-i\mathbf{k}\cdot(\mathbf{r}_{\alpha}+\mathbf{e}_{\mu}/2)}
\phi_{\lambda}(\mathbf{k})\nonumber\\
&=&
\frac{1}{\sqrt{N_s}}\sum_{\lambda =1,2, \, \mathbf{k} \in {\rm BZ}}\xi_{\lambda}(\mathbf{k})U^{\dagger}_{\mu\lambda}(\mathbf{k})
\nonumber\\
&\times&
e^{-i\mathbf{k}\cdot(\mathbf{r}_{\alpha}+\mathbf{e}_{\mu}/2)}
\phi_{\lambda}(\mathbf{k})
\, , \label{eq: curl lambda}
\quad \qquad 
\end{eqnarray}
where we have used the fact that columns of $U^{\dagger}_{\nu\lambda}(\mathbf{k})$ are eigenvectors of $Z_{\mu\nu}(\mathbf{k})$ with $\sum_{\nu}Z_{\mu\nu}(\mathbf{k})U^{\dagger}_{\nu\lambda}(\mathbf{k})=\xi_{\lambda}(\mathbf{k})U^{\dagger}_{\mu\lambda}(\mathbf{k})$, and that $\xi_{\lambda}(\mathbf{k})=0$ for $\lambda = 3,4$. Once again we find that the lattice curl depends only on the divergenceless modes $\lambda=1,2$. 

In order to study the Hamiltonian perturbatively in $1/s$ we then represent the $\lambda=1,2$ modes in terms of creation and annihilation operators, see Eq.~\eqref{diagonalisation}, and we do not need to consider the $\lambda=3,4$ modes any further.

\section{\label{app: normal ordering}
Normal Ordering the quartic part of the Hamiltonian
        }
We normal-order the terms in the quartic part of the Hamiltonian given in Eq.~\eqref{quartic}. Operators are normal ordered with respect to the creation and annihilation operators in which the quadratic Hamiltonian is diagonal. A string of operators is said to be normal ordered if all creation operators are on the left and all annihilation operators are on the right. 

We can always write any operator $A$ which is a linear superposition of creation and annihilation operators, e.g. $\phi_{ij}$ or $p_{ij}$ are such operators, as a sum of two parts: $A=A^++A^-$, where $A^+$ is a linear superposition of creation operators only and $A^-$ is a linear superposition of annihilation operators only. A contraction of two such operators $A$ and $B$ is defined as 
\begin{eqnarray}
\{A,B\}=\{A^++A^-,B^++B^-\}\equiv[A_-,B_+],
\end{eqnarray}
where the curly brackets signify a contraction, and the square brackets are commutators. From Eq.~\eqref{diagonalisation} in the main text, it then follows that 
\begin{eqnarray}
\{\phi_{\lambda}(\mathbf{k}),p_{\lambda^\prime}(\mathbf{q})\}
&=&
\frac{i}{2\tilde{s}}\delta_{\lambda,\lambda'}\delta_{\mathbf{k},-\mathbf{q}}\nonumber
\\
\{\phi_{\lambda}(\mathbf{k}),\phi_{\lambda^\prime}(\mathbf{q})\}
&=&
\frac{\omega(\mathbf{k})}{2\tilde{s}}\delta_{\lambda,\lambda'}\delta_{\mathbf{k},-\mathbf{q}}
\, ,
\nonumber\\
\{p_{\lambda}(\mathbf{k}),p_{\lambda^\prime}(\mathbf{q})\}
&=&
\frac{1}{2\tilde{s}\omega(\mathbf{k})}\delta_{\lambda,\lambda^\prime}\delta_{\mathbf{k},-\mathbf{q}}
\, , 
\label{eq: contractions phi p}
\end{eqnarray}
for $\lambda=1,2$. Because the contraction is a linear product of two operators, then $\{\alpha A + \beta B, C\}=\alpha\{A,C\}+\beta\{B,C\}$ (and it is anticommutative: $\{A,B\}=-\{B,A\}$). Contractions of operators which are linear superpositions of $\phi_{\lambda}(\mathbf{k})$ and $p_{\lambda}(\mathbf{k})$ can be computed straightforwardly as sums of the above contractions. 

To obtain the quartic contributions (see Eq.~\eqref{eq: quartic terms normal ordered} below), we need to evaluate two specific contractions, $\{ {\rm curl}_{\alpha \beta}\phi,{\rm curl}_{\alpha \beta}\phi \}$ and $\{ p_{ij}, p_{kl} \}$. 
The first contraction, using Eq.~\eqref{eq: curl lambda} and Eq.~\eqref{eq: contractions phi p}, gives 
\begin{eqnarray}
\{ {\rm curl}_{\alpha \beta}\phi,{\rm curl}_{\alpha \beta}\phi \}
=&& \nonumber\\
&&
\!\!\!\!\!\!\!\!\!\!\!\!\!\!\!\!
\!\!\!\!\!\!\!\!\!\!\!\!\!\!\!\!
= 
\frac{1}{N_s} \!\! \sum_{\mathbf{k},\mathbf{q}\in {\rm BZ}}
\sum_{\lambda,\lambda^\prime=1,2} \!\! 
U^{\dagger}_{\mu\lambda}(\mathbf{k})U^{\dagger}_{\mu\lambda^\prime}(\mathbf{q}) 
\nonumber\\
&& 
\!\!\!\!\!\!\!\!\!\!\!\!\!\!\!\!
\!\!\!\!\!\!\!\!\!\!\!\!\!\!\!\!
\times 
\xi_{\lambda}(\mathbf{k})\xi_{\lambda^\prime}(\mathbf{q})e^{- i \mathbf{k}(\mathbf{r}+\mathbf{e}_\mu/2) 
   - i \mathbf{q}(\mathbf{r}+\mathbf{e}_{\mu}/2)}
\nonumber\\
&&
\times\{\phi_{\lambda}(\mathbf{k}),\phi_{\lambda^\prime}(\mathbf{q})\}
\nonumber\\
&& 
\!\!\!\!\!\!\!\!\!\!\!\!\!\!\!\!
\!\!\!\!\!\!\!\!\!\!\!\!\!\!\!\!
= 
\frac{1}{2\tilde{s} N_s} 
\sum_{\mathbf{k}\in {\rm BZ}} \omega(\mathbf{k}) 
\sum_{\lambda=1,2} \xi^2_{\lambda}(\mathbf{k})
U^{\dagger}_{\mu\lambda}(\mathbf{k}) 
U^{\dagger}_{\mu\lambda}(-\mathbf{k})
\nonumber\\
&& 
\!\!\!\!\!\!\!\!\!\!\!\!\!\!\!\!
\!\!\!\!\!\!\!\!\!\!\!\!\!\!\!\!
= 
\frac{1}{2\tilde{s} N_s} 
\sum_{\mathbf{k}\in {\rm BZ}} \omega(\mathbf{k}) \: 
[Z^2(\mathbf{k})]_{\mu\mu}
\nonumber \\ 
&&
\!\!\!\!\!\!\!\!\!\!\!\!\!\!\!\!
\!\!\!\!\!\!\!\!\!\!\!\!\!\!\!\!
= 
\frac{z}{4 \tilde{s} N_s} 
\sum_{\mathbf{k}\in {\rm BZ}} \frac{1}{\omega(\mathbf{k})} 
\equiv \frac{C_0}{\tilde{s}} 
\, . 
\end{eqnarray}
(The choice of the same indices $\alpha \beta$ in both terms is intentional as it is the only term we will need). 
As before $\mathbf{e}_{\mu}=\mathbf{r}_{\beta}-\mathbf{r}_{\alpha}$, $z=6$ is the coordination number of a hexagonal plaquette, $C_0\approx 2.09$ and $[Z^2(\mathbf{k})]_{\mu\mu}$ is the $\mu\mu$ element of the square of the matrix $Z(\mathbf{k})$, after using the fact that $U^{\dagger}_{\mu\lambda}(-\mathbf{k})=U_{\lambda\mu}(\mathbf{k})$.
We evaluated $[Z^2(\mathbf{k})]_{\mu\mu}=\frac{z}{2\omega^2(\mathbf{k})}$ using the fact that $[Z^2(\mathbf{k})]_{\mu\mu}$ is independent of $\mu$, according for instance to the definition below Eq.~\eqref{eq: curl in k}. Therefore, $[Z^2(\mathbf{k})]_{\mu\mu} = \sum_\mu [Z^2(\mathbf{k})]_{\mu\mu} / 4$, which can be straightforwardly related to $\omega(\mathbf{k})=\sqrt{z}/|\xi_{\lambda = 1}(\mathbf{k})|$ via Eq.~\eqref{eq: xi}. 

In the second contraction, we only need to consider sites $ij$ and $kl$ belonging to the same plaquette centred on site $\alpha\beta \equiv \mathbf{r}+\frac{\mathbf{e}_{\sigma}}{2}$ of the dual pyrochlore lattice. The calculation proceeds in a similar manner, except that in this case there is a dependence on the relative displacement between the two sites: 
\begin{eqnarray}
&&\{p^{\mu}(\mathbf{r}+\frac{\mathbf{e}_{\sigma}}{2}+\mathbf{\Delta}_{\sigma \mu}),p^{\nu}(\mathbf{r}+\frac{\mathbf{e}_{\sigma}}{2}+\mathbf{\Delta}_{\sigma \nu})\}=
\nonumber\\
&&\quad 
= 
\frac{1}{N_s} \!\! \sum_{\mathbf{k},\mathbf{q}\in {\rm BZ}}
\sum_{\lambda,\lambda^\prime=1,2} \!\! 
U^{\dagger}_{\mu\lambda}(\mathbf{k})
U^{\dagger}_{\nu\lambda^\prime}(\mathbf{q}) 
\nonumber\\
&&\quad 
\times 
e^{- i \mathbf{k}(\mathbf{r}+\frac{\mathbf{e}_{\sigma}}{2}+\mathbf{\Delta}_{\sigma \mu}) 
   - i \mathbf{q}(\mathbf{r}+\frac{\mathbf{e}_{\sigma}}{2}+\mathbf{\Delta}_{\sigma \nu})}
\{p_{\lambda}(\mathbf{k}),p_{\lambda^\prime}(\mathbf{q})\}
\nonumber\\
&&\quad 
= 
\frac{1}{2 \tilde{s} N_s} \sum_{\lambda=1,2,\mathbf{k}\in {\rm BZ}}
U^{\dagger}_{\mu\lambda}(\mathbf{k})
U^{\dagger}_{\nu\lambda}(-\mathbf{k}) 
\frac{
  e^{- i \mathbf{k}(\mathbf{\Delta}_{\sigma \mu} - \mathbf{\Delta}_{\sigma \nu})}}{\omega(\mathbf{k})} 
\nonumber\\
&&\quad 
= 
\frac{1}{2 \tilde{s} N_s} 
\sum_{\mathbf{k}\in {\rm BZ}}
\frac{
  e^{- i \mathbf{k}(\mathbf{\Delta}_{\sigma \mu} - \mathbf{\Delta}_{\sigma \nu})}}{\omega(\mathbf{k})} 
\sum_{\lambda = 1,2}
U^{\dagger}_{\mu\lambda}(\mathbf{k})
U_{\lambda\nu}(\mathbf{k}) 
\nonumber\\
&&\quad 
= 
\frac{1}{2 z \tilde{s} N_s} 
\sum_{\mathbf{k}\in {\rm BZ}}
  e^{- i \mathbf{k}(\mathbf{\Delta}_{\sigma \mu} - \mathbf{\Delta}_{\sigma \nu})}
\omega(\mathbf{k}) 
\nonumber\\
&&\quad 
\times 
\sum_{\lambda = 1,2}
U^{\dagger}_{\mu\lambda}(\mathbf{k})
\xi^2_{\lambda}(k)
U_{\lambda\nu}(\mathbf{k}) 
\nonumber\\
&&\quad 
=
\frac{1}{2 z \tilde{s} N_s} 
\sum_{\mathbf{k}\in {\rm BZ}} 
e^{-i\mathbf{k}\cdot(\mathbf{\Delta}_{\sigma \mu}-\mathbf{\Delta}_{\sigma \nu})}
\omega(\mathbf{k})
\left[Z^2(\mathbf{k})\right]_{\mu \nu} 
\, .
\label{contraction}
\end{eqnarray}

The contraction of two $p$ operators on the same site follows
\begin{eqnarray}
\{p_{ij},p_{ij}\}=\frac{C_0}{z\tilde{s}}.
\end{eqnarray}

Using Wick's theorem we can express a string of operators as a normal ordering of that operator plus a sum over all possible pairwise contractions. The different terms in $\mathcal{H}_I$ can then be written as follows:
\begin{eqnarray}
\left[{\rm curl}_{\alpha \beta}\phi\right]^4
&=&
6\{{\rm curl}_{\alpha \beta}\phi,{\rm curl}_{\alpha \beta}\phi\}:\left[{\rm curl}_{\alpha \beta}\phi\right]^2:
\nonumber\\
&+&
3\{ {\rm curl}_{\alpha \beta}\phi,{\rm curl}_{\alpha \beta}\phi \}^2+:\left[{\rm curl}_{\alpha \beta}\phi\right]^4:
\nonumber
\end{eqnarray}
\begin{eqnarray}
p_{ij}^4
&=&
6\{p_{ij},p_{ij}\}:p_{ij}^2:+3\{p_{ij},p_{ij}\}^2 
\nonumber
\end{eqnarray}
\begin{eqnarray}
p_{ij}^2p_{kl}^2
&=&
:p_{ij}^2p_{kl}^2:+\{p_{ij},p_{ij}\}:p_{kl}^2:+\{p_{kl},p_{kl}\}:p_{ij}^2:
\nonumber\\
&+& 
2\{p_{ij},p_{kl}\}:p_{ij}p_{kl}:+\{p_{ij},p_{ij}\}^2+2\{p_{ij},p_{kl}\}^2
\nonumber
\end{eqnarray}
\begin{eqnarray}
\left[{\rm curl}_{\alpha \beta}\phi\right] p^2 \left[{\rm curl}_{\alpha \beta}\phi\right] 
&=& 
:\left[{\rm curl}_{\alpha \beta}\phi\right] p^2 \left[{\rm curl}_{\alpha \beta}\phi\right]:
\nonumber\\
&& 
\!\!\!\!\!\!\!\!\!\!\!\!\!\!\!\!\!\!\!\!
\!\!\!\!\!\!\!\!\!\!\!\!\!\!\!\!\!\!\!\!
\!\!\!\!\!\!\!\!\!\!\!\!\!\!\!\!\!\!\!\!
+  
:p^2:\{{\rm curl}_{\alpha \beta}\phi,{\rm curl}_{\alpha \beta}\phi\}+:\left[{\rm curl}_{\alpha \beta}\phi\right]^2:\{p,p\}
\nonumber\\
&& 
\!\!\!\!\!\!\!\!\!\!\!\!\!\!\!\!\!\!\!\!
\!\!\!\!\!\!\!\!\!\!\!\!\!\!\!\!\!\!\!\!
\!\!\!\!\!\!\!\!\!\!\!\!\!\!\!\!\!\!\!\!
+  
2 \{{\rm curl}_{\alpha \beta}\phi,p\}\{p,{\rm curl}_{\alpha \beta}\phi\}
\nonumber\\
&& 
\!\!\!\!\!\!\!\!\!\!\!\!\!\!\!\!\!\!\!\!
\!\!\!\!\!\!\!\!\!\!\!\!\!\!\!\!\!\!\!\!
\!\!\!\!\!\!\!\!\!\!\!\!\!\!\!\!\!\!\!\!
+  
\{{\rm curl}_{\alpha \beta}\phi,{\rm curl}_{\alpha \beta}\phi\}\{p,p\} 
\nonumber
\end{eqnarray}
\begin{eqnarray}
\left[{\rm curl}_{\alpha \beta}\phi\right]^2 p^2+p^2\left[{\rm curl}_{\alpha \beta}\phi\right]^2
&=&
:\left[{\rm curl}_{\alpha \beta}\phi\right]^2 p^2:
\nonumber\\
&& 
\!\!\!\!\!\!\!\!\!\!\!\!\!\!\!\!\!\!\!\!
\!\!\!\!\!\!\!\!\!\!\!\!\!\!\!\!\!\!\!\!
\!\!\!\!\!\!\!\!\!\!\!\!\!\!\!\!\!\!\!\!
\!\!\!\!\!\!\!\!\!\!\!\!\!\!\!\!\!\!\!\!
+  
:p^2\left[{\rm curl}_{\alpha \beta}\phi\right]^2:+2:p^2:\{{\rm curl}_{\alpha \beta}\phi,{\rm curl}_{\alpha \beta}\phi\}
\nonumber\\
&& 
\!\!\!\!\!\!\!\!\!\!\!\!\!\!\!\!\!\!\!\!
\!\!\!\!\!\!\!\!\!\!\!\!\!\!\!\!\!\!\!\!
\!\!\!\!\!\!\!\!\!\!\!\!\!\!\!\!\!\!\!\!
\!\!\!\!\!\!\!\!\!\!\!\!\!\!\!\!\!\!\!\!
+  
2:\left[{\rm curl}_{\alpha \beta}\phi\right]^2:\{p,p\}
\nonumber\\
&& 
\!\!\!\!\!\!\!\!\!\!\!\!\!\!\!\!\!\!\!\!
\!\!\!\!\!\!\!\!\!\!\!\!\!\!\!\!\!\!\!\!
\!\!\!\!\!\!\!\!\!\!\!\!\!\!\!\!\!\!\!\!
\!\!\!\!\!\!\!\!\!\!\!\!\!\!\!\!\!\!\!\!
+
2 \{{\rm curl}_{\alpha \beta}\phi,{\rm curl}_{\alpha \beta}\phi\}\{p,p\}
\nonumber\\
&& 
\!\!\!\!\!\!\!\!\!\!\!\!\!\!\!\!\!\!\!\!
\!\!\!\!\!\!\!\!\!\!\!\!\!\!\!\!\!\!\!\!
\!\!\!\!\!\!\!\!\!\!\!\!\!\!\!\!\!\!\!\!
\!\!\!\!\!\!\!\!\!\!\!\!\!\!\!\!\!\!\!\!
+  
4\{{\rm curl}_{\alpha \beta}\phi,p\}^2 
\label{eq: quartic terms normal ordered}
\end{eqnarray}
where :: denotes that the enclosed string of operators is normal ordered. 
Note that $\{{\rm curl}_{\alpha \beta}\phi,p\}\{p,{\rm curl}_{\alpha \beta}\phi\} = - \{{\rm curl}_{\alpha \beta}\phi,p\}^2$ and therefore the corresponding contributions in the last and second to last term above cancel exactly, and we do not need to compute them. 

Collecting all quadratic terms, e.g., $\{p_{ij},p_{ij}\}:p_{kl}^2:$, from the normal ordering of $\mathcal{H}_I$, we write down the Hartree correction to the quadratic Hamiltonian $\mathcal{H}_0$
\begin{eqnarray}
\!\!\!\!\!\!\!\!\!
\frac{\Delta \mathcal{H}_{0}}{g\tilde{s}^6} 
&=&
\left(\frac{-C_0}{\tilde{s}}\right)
\sum_{\alpha \beta}
:\left[{\rm curl}_{\alpha \beta}\phi\right]^2:
\label{correction}
\\
&&
+
\left(\frac{-3C_0}{\tilde{s}}\right) 
\sum_{ij} 
:p_{ij}^2:
\nonumber\\
&&
+\left(-\frac{1}{\tilde{s}}\right) \!\!\! 
\sum_{\lambda,\lambda' = 1,2, \, \mathbf{k}\in{\rm BZ}} \!\! 
C_{\lambda \lambda '}(\mathbf{k})
:p_{\lambda}(\mathbf{k}) p_{\lambda '}(-\mathbf{k}): 
\nonumber\\
&=&
-\frac{C_0}{\tilde{s}} \!\!\! 
\sum_{\lambda,\lambda' = 1,2, \, \mathbf{k}\in{\rm BZ}}
\Big[
\frac{C_{\lambda \lambda '}(\mathbf{k})}{C_0}
:p_{\lambda}(\mathbf{k})p_{\lambda'}(-\mathbf{k}):
\nonumber\\
&&
+\delta_{\lambda\lambda'}\xi_{\lambda}^2(\mathbf{k})
:|\phi_{\lambda}(\mathbf{k})|^2:
+3\delta_{\lambda\lambda'}
:|p_{\lambda}(\mathbf{k})|^2:
\Big]
\nonumber 
\end{eqnarray}
where the matrix
\begin{eqnarray}
&&C_{\lambda \lambda '}(\mathbf{k}) 
\equiv 
\frac{1}{z N_s}
\sum_{\mathbf{q} \in {\rm BZ}} \; 
\sum_{\sigma,(\mu, \nu \neq \sigma)}
  \omega(\mathbf{q})\left[ Z^2(\mathbf{q}) \right]_{\mu \nu}
\\
&&\; \times 
\cos\left[(\mathbf{k}+\mathbf{q})\cdot\mathbf{\Delta}_{\sigma \mu} \right]
\cos\left[(\mathbf{k}+\mathbf{q})\cdot\mathbf{\Delta}_{\sigma \nu} \right]
U^{\dagger}_{\mu \lambda}(\mathbf{k}) 
U^{\dagger}_{\nu \lambda '}(-\mathbf{k}).
\nonumber 
\end{eqnarray}
The term in Eq.~\eqref{correction} that mixes different photon polarisations arises from contractions of $p$ operators on different lattice sites
\begin{eqnarray}
&&
\!\!\!\!\!\!\!\!\!\!\!\!\!
\sum_{\alpha\beta}\sum_{ij,kl\in\alpha\beta} 
\{p_{ij}p_{kl}\} : p_{ij}p_{kl} : \; = 
\nonumber\\
&&
=\sum_{\mathbf{r}\in{\rm fcc}}
\sum_{\sigma,\mu\neq\sigma,\nu\neq\sigma}
\sum_{a,b=\pm 1}
\nonumber\\
&&\times\{p^{\mu}(\mathbf{r}+\frac{\mathbf{e}_{\sigma}}{2}+a\mathbf{\Delta}_{\sigma \mu}),p^{\nu}(\mathbf{r}+\frac{\mathbf{e}_{\sigma}}{2}+b\mathbf{\Delta}_{\sigma \nu})\}\nonumber\\
&&\times :p^{\mu}(\mathbf{r}+\frac{\mathbf{e}_{\sigma}}{2}+a\mathbf{\Delta}_{\sigma \mu})p^{\nu}(\mathbf{r}+\frac{\mathbf{e}_{\sigma}}{2}+b\mathbf{\Delta}_{\sigma \nu}):.
\end{eqnarray}
Substituting in for the contraction using Eq.~\eqref{contraction} and transforming to the eigenbasis of $\mathcal{H}_0$ gives us the expression in Eq.~\eqref{correction}.

To obtain the single-particle spectrum correct to order $s^{-1}$, first-order perturbation theory requires us to diagonalise the matrix 
\begin{eqnarray}
H_{\lambda\lambda'}(\mathbf{k})
&=&
\langle {\rm g.s.}|a_{\lambda}(\mathbf{k})\left(\mathcal{H}_0+\Delta \mathcal{H}_{0}\right) a^{\dagger}_{\lambda'}(\mathbf{k})|{\rm g.s.}\rangle 
\\
&=&
\frac{2zg\tilde{s}^5}{\omega(\mathbf{k})}
\Big[
  \delta_{\lambda\lambda'}\left(
	  1-\frac{C_0}{2\tilde{s}}-\frac{3C_0}{2\tilde{s}z}
	\right)
	-\frac{1}{2\tilde{s}z}C_{\lambda\lambda'}(\mathbf{k})
\Big]
\, , 
\nonumber
\end{eqnarray}
obtained from Eq.~\eqref{diagonalisation}, Eq.~\eqref{eq: H0} (without the constant $1/2$ term), and Eq.~\eqref{correction}. 
The single-particle energy spectrum is then given by its eigenvalues:
\begin{eqnarray}
&&
\epsilon_{\eta}(\mathbf{k}) = 
\nonumber\\
&&
=
\frac{2zg\tilde{s}^5}{\omega(\mathbf{k})}\left[
  1-\frac{1}{2\tilde{s}}\left( 
	  C_0
		+
		\frac{3C_0}{z}
		+
		\frac{1}{z}C_{11}(\mathbf{k})
		\pm \frac{1}{z}|C_{12}(\mathbf{k})| 
	\right) 
\right]
\nonumber\\
&&
=
\frac{2zgs^5}{\omega(\mathbf{k})}\Big[
  \frac{1}{2s}\left(
    5 - C_0 - \frac{3C_0}{z} - \frac{1}{z}C_{11}(\mathbf{k})
		\mp\frac{1}{z}|C_{12}(\mathbf{k})|
	\right) 
\nonumber\\
&&
+ 1 +\mathcal{O}(s^{-2})
\Big]
\end{eqnarray}
where the plus or minus sign is chosen for the eigenvalues $\epsilon_{\eta=1}(\mathbf{k})$ and $\epsilon_{\eta=2}(\mathbf{k})$ respectively. Note that $C_{12}(\mathbf{k}) = C_{21}(-\mathbf{k}) = C^*_{21}(\mathbf{k})$, from the properties of the matrix $U_{\mu\nu}(\mathbf{k})$. 
Moreover, $C_{11}(\mathbf{k}) = C_{22}(\mathbf{k})$, which follows from the fact that $\left[Z^2(\mathbf{k})\right]_{\mu\nu}$ is a symmetric matrix and from the relation $U^{\dagger}_{\nu 1}(-\mathbf{k})=U^{\dagger}_{\nu 2}(\mathbf{k})$, demonstrated by the following eigenvalue equation
\begin{eqnarray}
\sum_{\nu}Z_{\mu\nu}(\mathbf{k})U^{\dagger}_{\nu 1}(-\mathbf{k})
&=&
-\sum_{\nu}Z_{\mu\nu}(-\mathbf{k})U^{\dagger}_{\nu 1}(-\mathbf{k})
\nonumber\\
&=&
-\xi_1(-\mathbf{k})U^{\dagger}_{\mu 1}(-\mathbf{k})
\nonumber\\
&=&
\xi_2(\mathbf{k})U^{\dagger}_{\mu 1}(-\mathbf{k}) 
\, . 
\end{eqnarray}

The splitting $|C_{12}(\mathbf{k})|$ vanishes at $\mathbf{k}=\mathbf{0}$  and the speed of light is given by
\begin{eqnarray}
\frac{c}{gs^5}
&=& 
\frac{2z}{\omega(\mathbf{k})}\left[
  1+\frac{1}{2s}\left(
    5
		-C_0
    -\frac{3C_0}{z}
    -\frac{1}{2z}C_{11}(\mathbf{0})
  \right)
\right]
\nonumber \\ 
&+&
\mathcal{O}\left(s^{-2}\right)
\nonumber\\
&=&
\frac{2z}{\omega(\mathbf{k})}\left(1+\frac{0.846}{s}\right)
\, , 
\end{eqnarray}
where we used the fact that $C_{11}(\mathbf{k})$ is linear in $\mathbf{k}$ near $\mathbf{k} = \mathbf{0}$. 

Collecting the constant terms 
from the normal ordering of $\mathcal{H}_{I}$ we obtain the Hartree 
correction to the ground state energy 
\begin{eqnarray}
\langle {\rm g.s.}|\mathcal{H}_I |{\rm g.s.}\rangle 
&=& 
-g \tilde{s}^6 \sum_{\alpha\beta} \left\{ \vphantom{\sum_{ij \in \alpha\beta}}
\frac{1}{4} \{ {\rm curl}_{\alpha \beta}\phi,{\rm curl}_{\alpha \beta}\phi \}^2 
\right. 
\nonumber \\ 
&-& \frac{3}{2} \sum_{ij \in \alpha\beta} \{p_{ij},p_{ij}\}^2 
\nonumber \\ 
&+&\frac{1}{4} \sum_{ij \in \alpha\beta}
  \{{\rm curl}_{\alpha \beta}\phi,{\rm curl}_{\alpha \beta}\phi\}\{p,p\}
\nonumber \\ 
&+& \frac{1}{4} \sum_{ij \in \alpha\beta}
  \{{\rm curl}_{\alpha \beta}\phi,{\rm curl}_{\alpha \beta}\phi\}\{p,p\} 
\nonumber \\ 
&+&\left.\frac{1}{4} \sum_{ij,kl \in \alpha\beta} 
  \left[ \{p_{ij},p_{ij}\}^2+2\{p_{ij},p_{kl}\}^2 \right] 
\right\} 
\nonumber \\ 
&=& 
-g \tilde{s}^4 (C_0)^2 
\sum_{\alpha\beta} \left\{ \vphantom{\sum_{ij \in \alpha\beta}}
  \frac{1}{4}-\frac{3}{2z}+\frac{1}{4}+\frac{1}{4}+\frac{1}{4} 
\right\} 
\nonumber \\ 
&+& 
- \frac{g \tilde{s}^6}{2} \sum_{\alpha\beta} 
  \sum_{ij,kl \in \alpha\beta} \{p_{ij},p_{kl}\}^2 
\nonumber \\ 
&=&
-g\tilde{s}^4N_p\left(\frac{3}{4}C_0^2+C_1\right) 
\, , 
\end{eqnarray} 
where $C_1\approx 0.468$ follows from performing the summation over $\alpha\beta$ 
and over $ij$ ($kl$) on the square of Eq.~\eqref{contraction} and $N_p=4N_s$ is the total number of plaquettes. Treating $\mathcal{H}_I$ as a perturbation, to first order (Hartree-Fock) the ground-state energy is given by
\begin{eqnarray}
\frac{E}{N_pg\tilde{s}^6}&=&-2+
\frac{\sqrt{z}}{N_p\tilde{s}}\sum_{\mathbf{k}\in{\rm BZ},\lambda}|\xi_{\lambda}(\mathbf{k})|\nonumber\\
&&+\frac{1}{N_p}\langle{\rm g.s.}|\frac{\mathcal{H}_I}{g\tilde{s}^6}|{\rm g.s.}\rangle+\mathcal{O}(\tilde{s}^{-4}).
\end{eqnarray}
Expanding in $s^{-1}$, we can now write the ground state energy correct to order $s^{-2}$ 
\begin{eqnarray}
&&\frac{E}{N_pgs^6}=-2\frac{\tilde{s}^6}{s^6}+2C_0\frac{\tilde{s}^5}{s^6}-\left(\frac{3}{4}C_0^2+C_1\right)\frac{\tilde{s}^4}{s^6}+\mathcal{O}\left(\frac{\tilde{s}^2}{s^6}\right)\nonumber\\
&&=
-2+\frac{A_1}{s}+\frac{A_2}{s^2}+\mathcal{O}(s^{-3}),
\end{eqnarray}
where $A_1=2C_0-6\approx -1.820$ and $A_2=5C_0-\frac{15}{2}-\frac{3}{4}C_0^2-C_1\approx -0.793$.

\section{\label{app: gMFT parameter}
Fluctuations of the gauge mean-field
        }
The fluctuations of the zero-energy modes of $\phi_{ij}$, which are proportional to $ \chi_j - \chi_i$, i.e., $\phi_{\lambda=3} (\mathbf{k})$ and $\phi_{\lambda =4} (\mathbf{k})$, do not contribute to the gauge-independent expectation values of the electric field ${\rm curl_{\alpha\beta}}\phi$, nor to the dynamical, transverse part of the magnetic field $S^z_{ij}$ (the longitudinal part is set to zero by the ice rules). The physical ground state wavefunction (in $\phi_{ij}$ space) is a function of only the transverse modes $\phi_{\lambda=1} (\mathbf{k})$ and $\phi_{\lambda =2} (\mathbf{k})$, see Eq.~\eqref{gs}, and can be multiplied by any function of the longitudinal modes $F \left[ \phi_{\lambda=3} (\mathbf{k}), \phi_{\lambda =4} (\mathbf{k}) \right]$, without altering the expectation values of gauge-independent observables  -- this is the quantum analogue of classical gauge fixing. Each gauge corresponds to a particular choice of the function $F \left[ \phi_{\lambda=3} (\mathbf{k}), \phi_{\lambda =4} (\mathbf{k}) \right]$. It is interesting to consider a gauge, where the gauge field
\begin{eqnarray}
\langle S^x \rangle 
\equiv 
\frac{1}{2N_p}\sum_{ij} \langle S_{ij}^+ + S_{ij}^- \rangle
\label{eq: expectation}
\end{eqnarray}
has a non-zero expectation value, i.e., it has long-range order and spontaneously broken symmetry.  There are many choices of $F \left[ \phi_{\lambda=3} (\mathbf{k}), \phi_{\lambda =4} (\mathbf{k}) \right]$, which give a non-zero expectation value of the gauge field. We shall make the choice that minimises the fluctuations of the gauge field order parameter 
$\langle S^x \rangle$
to first order in $1/s$. This can be referred to as the maximally coherent gauge and essentially removes the effect of longitudinal fluctuations on the order parameter, so that the only fluctuations that remain are transverse. The relative magnitude of this reduction can then be interpreted as the size of the corrections to the RK state, see Eq.~\eqref{gs}, which is the exact ground state in the $s=\infty$ limit, contains no transverse fluctuations, and is continuosly connected to the ground state at finite $s$. Our semi-classical expansion corresponds to a perturbation of the RK state, and small corrections would support the RG arguments in favour of the stability of the $s=\infty$ fixed point and the deconfining phase connected to it.  Note that the Monte Carlo calculations of Ref.~\onlinecite{Shannon} also analysed the adiabatic continuity to the RK state, but via a chemical potential term rather than a semi-classical expansion.

Furthermore, the expectation value in Eq.~\eqref{eq: expectation} in the maximally coherent gauge can be interpreted as the gauge mean-field order parameter of Ref.~\onlinecite{Savary1}, i.e., the expectation value of the operator
\begin{eqnarray}
  e^{i \psi_i} S_{ij}^+ e^{-i \psi_j} ,
\end{eqnarray} 
where $\psi$ is the phase of the slave-boson field, and $e^{\pm i \psi_{i,j}}$ creates a magnetic monopole/antimonopole at site $i,j$. Through a gauge-fixing procedure, e.g., {\it \`{a} la} Feddeev-Popov (neglecting periodic boundary conditions on $\phi_{ij}$), the unbounded fluctuations in the longitudinal component ($\lambda=3,4$) of the spin phase $\phi_{ij}$, proportional to $\chi_j - \chi_i$, can be offset by the fluctuations of the phase of the slave boson field $\psi$, so that the above expectation value is non-zero. This motivates the mean-field decoupling of Ref.~\onlinecite{Savary1}. The expectation value of the above operator with respect to the true ground state is then equivalent to the expectation value of  $S_{ij}^+$ in the ground state with modified longitudinal fluctuations of $\phi_{ij}$. 

As outlined before, the modification involves multiplying the ground state, which is a function of $\phi_{\lambda=1,2}(\mathbf{k})$ only, by a function of the longitudinal modes $F \left[ \phi_{\lambda=3} (\mathbf{k}), \phi_{\lambda =4} (\mathbf{k}) \right]$. The reduction of the gauge mean-field from its maximum value of $\frac{1}{2}$ becomes a measure of the remaining transverse fluctuations in $\phi_{ij}$ (which cannot be absorbed by the slave boson field). Notice that 
$\langle S^x \rangle \neq 0$ 
(long-range order) in the deconfining phase, whereas 
$\langle S^x \rangle =0 $ 
(absence of long-range order) would be indicative of the confining phase.

Using our semi-classical approach, we can calculate 
$\langle S^x \rangle$ in the maximally coherent gauge
to first order in zero point fluctuations, i.e., to first order in $1/s$, 
\begin{eqnarray}
\frac{\langle S^x \rangle}{s} 
&=&
\frac{\tilde{s}}{2 s N_p} \sum_{ij} \langle 2- \phi_{ij}^2 - p_{ij}^2\rangle
+ \mathcal{O} (\tilde{s}^{-2})
\nonumber\\
&=&
\frac{\tilde{s}}{ s} -
\frac{\tilde{s}}{2 s N_p}  
\sum_{\mathbf{k} \neq \mathbf{0}, \lambda=1,2} \langle |\phi_{\lambda} (\mathbf{k})|^2 
+|p_{\lambda} (\mathbf{k})|^2
\rangle_{\rm g.s.}
\nonumber\\
&-&
 \frac{\tilde{s}}{2 s N_p} 
\sum_{\mathbf{k} \neq \mathbf{0}, \lambda=3,4} \langle |\phi_{\lambda} (\mathbf{k})|^2 
+|p_{\lambda} (\mathbf{k})|^2
\rangle_{F}
+ \mathcal{O} (\tilde{s}^{-2})
\nonumber\\
&=&
\frac{\tilde{s}}{ s} 
- \frac{1}{2 s N_p} 
\sum_{\mathbf{k} \neq \mathbf{0}} 
\left[  \omega(\mathbf{k}) + \frac{1}{\omega(\mathbf{k})}  \right] -\frac{1}{4s}
+ \mathcal{O} (\tilde{s}^{-2})
\nonumber\\
&=& 1 + \frac{1}{4s} - \frac{1}{2sN_p} \sum_{\mathbf{k} \neq \mathbf{0}} 
\left[  \omega(\mathbf{k}) + \frac{1}{\omega(\mathbf{k})}  \right]
+ \mathcal{O} (s^{-2}),
\nonumber\\
&=& 1- \frac{D}{s}  + \mathcal{O} (s^{-2}),
\end{eqnarray} 
where the the expectation value $\langle\rangle_{\rm g.s.}$ is taken with respect to the physical ground state, i.e., the bosonic vacuum defined in Eq.~\eqref{diagonalisation}. The expectation value $\langle\rangle_{\rm F}$ is taken with respect to such wavefunction $F \left[ \phi_{\lambda=3} (\mathbf{k}), \phi_{\lambda =4} (\mathbf{k}) \right]$ which minimises it: $\langle |\phi_{\lambda} (\mathbf{k})|^2  
+|p_{\lambda} (\mathbf{k})|^2
\rangle_{F} = 1/\tilde{s}$ is the minimum expectation value for each mode ($\lambda=3,4$, $\forall \: \mathbf{k}$) and corresponds to the simple harmonic oscillator ground state energy. (To see this one can use Eq.~\eqref{diagonalisation} with the choice $\omega (\mathbf{k})=1$ to express  $|\phi_{\lambda} (\mathbf{k})|^2  
+|p_{\lambda} (\mathbf{k})|^2$ in terms of creation and annihilation operators -- the state $F$ is then the vacuum of these bosonic operators.) We have also excluded the global zero-mode $\mathbf{k}= \mathbf{0}$ from the sums, and calculated $D=0.019$. We find
\begin{eqnarray}
 \langle S^x \rangle = 0.481 \qquad {\rm for} \qquad s=\frac{1}{2}.
\end{eqnarray}

The mean-field theory of Ref.~\onlinecite{Savary1} gives $\langle S^x \rangle =\frac{1}{2}$ in the region of parameter space where the ring-exchange model is applicable. We thus find that zero-point fluctuations only give a small $\sim 4\%$ correction to mean-field theory. In our semi-classical expansion, this small reduction can be interpreted as a smal correction to the RK state and supports the other arguments we have presented for the stability of the classical $s=\infty$ fixed point and the deconfining phase.



\end{document}